\documentclass[10pt, preprint, longbibliography]{revtex4-1}
\usepackage{amsmath}

\begin{document}

\title{Canonical (and non-canonical) Variables: A Differential Approach}
\author{Stephen D. Webb}
\email{swebb@txcorp.com}
\affiliation{Tech-X Corporation, 5621 Arapahoe Ave. Suite A, Boulder, Colorado 80303, USA}

\begin{abstract}
The traditional method of teaching canonical transformations involves the introduction of generating functions of various types. This method obscures the underlying structure of the Hamiltonian least-action principle, and can make a straightforward concept seem arcane. In this article, I present a method for calculating canonical changes of variable in Hamiltonian mechanics using a differential approach which is much more straightforward. This method handles canonical variables directly, but also returns the correct equations of motion for non-canonical variables. It is also much more algebraic than generating functions, making it easier to present in a systematic manner.
\end{abstract}

\maketitle

\section{Introduction}

Canonical transformations are the strongest tool presented by Hamiltonian mechanics over Lagrangian mechanics, specifically transformations that yield the action-angle variables and carry conserved quantities. There are many equivalent definitions of canonical transformations of variables, but they all boil down to the statement that a set of variables $(P, Q)$ are canonical transforms of the variables $(p, q)$ if they preserve the symplectic structure of the equations of motion.

The most commonly presented method for obtaining canonical transformation are the generating functions outlined in many classical mechanics textbooks (\cite{GPS, CS, lan_cla, jos_sal} to name a few). Derivatives of these generating functions yield the new set of canonical variables in terms of the old set, and in this sense they are straightforward. Most classical mechanics textbooks emphasize obtaining the generating function, whence they go about carrying out canonical transformations this way.

In this paper, I present a method of carrying out canonical transformations which is equivalent to the generating function method, but which is more flexible in the sense that one does not \emph{need} the generating function to carry out a canonical transformation. This has the advantage that new canonical variables can be obtained more or less by algebraic methods, and so new canonical variables may be introduced readily to any Hamiltonian and any new set of variables, regardless of whether a generating function is known. It furthermore has the advantage of generalizing to non-canonical coordinates if one should ever need these things.

It is important to stress that no new physics derives from any choice of coordinate system. The use of canonical variables is merely a convenience, although a very useful convenience, in describing dynamical systems.

In Section II, I present a concise derivation of the method that springs from the Legendre transformation from Lagrangian to Hamiltonian action integrals. The general rules for carrying out canonical transformations using this new method are presented in Section III. Section IV presents a series of applications and academic problems which use this method. These range from the change of independent variable to a use of complex-valued canonical variables to cope with the two dimensional harmonic oscillator potential. Section V presents the Dirac ``raising" and ``lowering" operators from quantum mechanics as an example of non-canonical variable transformations in classical mechanics. 

\section{Lagrangian to Hamiltonian Action Integrals}

The Lagrangian action integral of classical mechanics, with independent variable $s$, is given by
\begin{equation}
S = \int_{s_0}^{s_f} ~ L(q, \dot{q}, s) ~ ds
\end{equation}
where overdot denotes a derivative with respect to $s$. The equations of motion are determined by the trajectories in configuration space for which the action integral is minimized, viz. $\delta S = 0$. I will assume the reader is familiar with the calculus of variations enough that this first variation statement is clear. If not, there is an excellent general reference by Weinstock \cite{wein} on the calculus of variations.

Now, the Hamiltonian is a Legendre transformation on the Lagrangian. The canonical momentum is defined by
\begin{equation}
p_\imath = \frac{\partial L}{\partial \dot{q}_\imath}
\end{equation}
for whichever component is desired. The Hamiltonian is then defined by the transformation
\begin{equation}
H = p_\imath \dot{q}^\imath - L(q, p, s)
\end{equation}
Inserting this new definition into the action integral above gives
\begin{equation} \label{Hamiltonianaction}
S = \int \left ( p_\imath \dot{q}^{\imath} - H(p, q, s) \right )~ ds
\end{equation}
Hamilton's equations are a direct consequence of minimizing this integral with respect to $(p, q)$.

At this point I am taking liberties with the rigors of differential forms, but for our purposes we will not get into too much trouble. Notice that $\dot{q} ~ds = d q$, so now the action integral is given by
\begin{equation}\label{diffaction}
S = \int p_\imath dq^{\imath} - H(p, q, s) ds
\end{equation}
where the implied summation is used. This is the action integral of Hamiltonian mechanics. The previous version works fine if we want $s$ to be the independent variable, but this differential form will be the focus of the remainder of this note. It is instructive to apply the least action principle to equation (\ref{Hamiltonianaction}) to see that this does indeed return the familiar Hamilton equations of motion.

\section{Canonical Transformations \& Generating Functions}

The method presented here recasts canonical transformations from a method of generating functions to exploiting the differential form of the Hamiltonian action integral. The differential form $p_\imath d q^{\imath} - H ds$ contains all the information necessary to carry out canonical (and non-canonical) changes of variable.

A set of variables $(p, q)$ are canonical if their action integral is in the differential form in equation (\ref{diffaction}). Because of the way we constructed the action integral from the Lagrangian, we know this statement to be true because we have used the definition of the canonical momentum to reach this point.

Suppose the Lagrangian is expressed in the generalized coordinate sets such that
\begin{equation}
L'(Q, \dot{Q}, s') = L(q, \dot{q}, s)
\end{equation}
Such a statement clearly models the same dynamical system. The action integral for each one is given by
\begin{equation}
\int L'(Q, \dot{Q}, s') ds' = \int L(q, \dot{q}, s) ds
\end{equation}
The Hamiltonian action integrals are then given by
\begin{equation}
\int \left (P \dot{Q} - H' \right ) ds' = \int \left ( p \dot{q} - H \right ) ds
\end{equation}
The differential form is then given by
\begin{equation}
\int \left (P ~dQ - H' ds' \right ) = \int \left ( p ~ dq - H ds \right )
\end{equation}
Both integrals lead to Hamiltonian equations of motion, therefore $(P,Q)$ and $(p,q)$ are both canonical variables. By construction, these Hamiltonians correspond to the same physical system, and therefore the differentials must be equal up to an exact differential which vanishes upon first variation. Therefore, canonical changes of variable preserve the differential form
\begin{equation}\label{canonical}
\left ( P ~dQ - H' ds' \right ) = \left ( p ~ dq - H ds \right ) ~ + ~ \left ( \frac{d F}{d s} ds \right )
\end{equation}
where $F$ is the generating function for the canonical transformation just carried out. The conventional approach from here would be to place a heavy emphasis on all the ways $F$ can be used. This approach has its uses, but is frequently overemphasized to the point that many professional physicists rely entirely on generating functions as the characterizing feature of canonical transformations.

An instructor interested in a broader approach should strongly consider looking at emphasizing the full differential relation in equation (\ref{canonical}), and not overemphasize the importance of the generating function. The advantage granted here is that one does not need to look up a generating function and calculate old variables in terms of new variables and carry out an inversion. This utility is perhaps best demonstrated with a series of examples.

\section{Examples}

As canonical transformations are a practical tools in classical mechanics, their uses are best illustrated by applications. Therefore, it is now prudent to consider a few examples of how this method works in practice. The goal of this section is to provide a number of examples of increasing sophistication of using this more direct method, bypassing the generating function entirely. Several of these examples are well-known and handled with generating functions in standard textbooks. The section on harmonic potentials, to the knowledge of the author, cannot be found anywhere else.

\subsection{Changing Independent Variable}

Suppose we wish to make one of the $q$ variables the independent variable. This is simple enough. We identify $q^* = s'$. Immediately, we find that $H' = -p^*$, and that $(q^*)' = s$ and $(p^*)' = - H$ by simple inspection of the differential form.

The benefit here is obvious and intuitive. In the Lie algebraic language of quantum mechanics the momentum is the hermitian operator which generates the translation Lie group -- $p_x$ is the generator of $x$ translations. That is made clear here: since $x$ is the new independent variable, and we want to translate our other dynamical variables through $x$, the generator of those translations is the new ``Hamiltonian" $p_x$. Thus, this formulation keeps Hamiltonian mechanics close to its modern formulation of Lie groups (see, e.g., \cite{dragt, lich_lieb}). As an exercise, show that
\begin{equation}
\frac{d f}{d x} = \frac{\partial f}{\partial x} + \{ f, p_x \}
\end{equation}
where $\{ A, B \}$ is the Poisson bracket. This is equivalent to the above statement of $p_x$ generating translations in the $x$-direction.

\subsection{Constant Velocity Transformation}

In accelerators, the chosen Frenet-Serret coordinates are chosen around a \emph{design trajectory} (see, e.g., \cite{leeacc}) for a particle moving at exactly the design energy and exactly where the trajectories are intended. In this context, the useful longitudinal coordinates are those which deviate from the design particle trajectory.

Suppose we want to make a new longitudinal variable $\psi = s - x/v$ the new independent variable, and let $x' = x + s/v$ describe variation from a given longitudinal trajectory. Such a coordinate system is of practical importance to many accelerator physics applications. I consider only two dimensions here, as the motion transverse to $v$ is left unaffected by this transformation to a moving frame.

Start with
\begin{subequations}
\begin{equation}
d \psi = ds - dx/v = ds'
\end{equation}
\begin{equation}
dx' = dx - v ds
\end{equation}
\end{subequations}
By direct insertion, this gives the differential form being required to satisfy the constraint
\begin{equation}
p dx - H ds = p' dx' - H' d \psi = p' (dx - v ds) - H' (ds - dx/v)
\end{equation}
Then we must satisfy the equations
\begin{subequations}
\begin{equation}
p = p' + H'/v
\end{equation}
\begin{equation}
- H = - p' v - H'
\end{equation}
\end{subequations}
Direct inversion of these equations gives
\begin{equation}
\left(
\begin{array}{c}
p'   \\
 -H' 
\end{array}
\right)
= 
\left(
\begin{array}{c}
p + H/v   \\
p v - H
\end{array}
\right)
\end{equation}
Therefore
\begin{equation}
H' = H(p', x'; \psi) - p' v
\end{equation}
and the equations of motion follow from this transformation, and are symplectic by construction. It is also an interesting and straightforward exercise to watch how a time rescaling propagates through the action integral.

\subsection{The Harmonic Oscillator and Action-Angle Variables}

A frequent example of canonical transformations are the action-angle variables, and the most common first example of that is the harmonic oscillator. This is usually a student's first introduction to generating functions, canonical transformations, and all the formalism that follows. It would be nice to use this example as a swift introduction to the utility of this differential approach.

The objective of action-angle variables is to write the Hamiltonian as independent of the canonical coordinate. Let us then consider the simple harmonic oscillator
\begin{equation}
H = \frac{1}{2} \left ( p^2 + \omega_0^2 q^2 \right )
\end{equation}
In the differential form, we want to find a set of coordinates which satisfy
\begin{equation}
p dq - H dt = J d \theta - H'(J) dt
\end{equation}
where it would be nice (although not generally necessary) to keep the time variable unchanged. For a conserved system, given any initial energy $E_0$, the particle will trace out the ellipse defined by
\begin{equation}
E_0 = \frac{1}{2} \left ( p^2 + \omega_0^2 q^2 \right )
\end{equation}
We also know that it traces out these ellipses at a fixed frequency. We may therefore consider rewriting the coordinates as
\begin{subequations}
\begin{equation}
p = A \sqrt{J} \sin \theta
\end{equation}
\begin{equation}
x = B \sqrt{J} \cos \theta
\end{equation}
\end{subequations}
where $\theta$ parameterizes the point on the ellipse and $\sqrt{J}$ parameterizes the amplitude. Inserting this into the old Hamiltonian gives
\begin{equation}
H' = \left ( A^2 + \omega_0^2 B^2 \right ) J
\end{equation}
and the differential takes the form
\begin{equation}
A \sqrt{J} \sin \theta ~ d \left ( B \sqrt{J} \cos \theta \right ) = J d \theta
\end{equation}
Expanding the differential and looking at the left-hand side, we find
\begin{equation}
\begin{split}
A B \left (\frac{1}{4} \sin 2 \theta d J + \frac{1}{2} (1 + \cos 2 \theta ) J d \theta \right )= \\ A B \left ( d [J \sin 2 \theta] + \frac{1}{2} J d \theta \right )
\end{split}
\end{equation}
The first term is an exact differential (in fact, it is a representation of the generating function for this canonical transformation) and does not contribute to the equations of motion. The second term requires that $A B = 2$ to match with the $J d \theta$ term above.

Looking at the transformed Hamiltonian, it would be most convenient if
\begin{subequations}
\begin{equation}
A = \sqrt{2 \omega_0}
\end{equation}
\begin{equation}
B = \sqrt{\frac{2}{\omega_0}}
\end{equation}
\end{subequations}
We have thus obtained the action-angle variables and their corresponding Hamiltonian for the harmonic oscillator, and the generating function for the transformation as a byproduct.

\subsection{Complex Variables and the Two Dimensional Harmonic Oscillator}

As an exercise in this method that is perhaps a bit more interesting, consider a double harmonic oscillator with identical frequencies in the $x$ and $y$ directions. This example is worked out using generating functions in the underused Corben and Stehle \cite{CS}. The Hamiltonian for this system is given by
\begin{equation}
H = \frac{1}{2} \left ( p_x^2 + p_y^2 \right ) + \frac{1}{2} \omega_0^2 \left ( x^2 + y^2 \right )
\end{equation}
it might be convenient (and useful) to express this in terms of complex variables, since the potential may be written as $\omega_0^2 z z^*$ where $z = x + \imath y$. This requires that we find the canonical momenta conjugate to these complex valued variables.

It would be very difficult in general to divine the correct generating function for these variables. Using the differential approach, however, it is simply a matter of going through the process. $x = (z + z^*)/2$ and $y = (z- z^*)/(2 \imath)$. Therefore the differential form is given by
\begin{equation}
\begin{split}
p_x~ (dz + dz^*)/2 - \imath p_y~ (dz - dz^*)/2 - H ds = \\ p_1 ~dz + p_2 ~ dz^* - H' ds
\end{split}
\end{equation}
Pairing off coordinates and again follow through with the algebra we find that
\begin{subequations}
\begin{equation}
p^1 = p = (p_x - \imath p_y)/2
\end{equation}
\begin{equation}
p^2 = p^* = (p_x + \imath p_y)/2
\end{equation}
\end{subequations}
and then, by direct substitution, we get that the new Hamiltonian is given by
\begin{equation}
H' = 2 p p^* + \frac{1}{2} \omega_0^2 z z^*
\end{equation}
which, by a simple rescaling by $\sqrt{2}$ (note that in this formulation it is straightforward to see that $p' = a p$ and $x' = x/a$ are canonically conjugate) gives the Hamiltonian as
\begin{equation}
H' = p p^* + \omega_0^2 z z^*
\end{equation}
The equations of motion can be immediately determined. As another exercise, determining the action-angle variables in this new coordinate system is useful for perturbation theory. Try letting
\begin{subequations}
\begin{equation}
p = \sqrt{ \imath J \omega_0} e^{\imath \theta}
\end{equation}
\begin{equation}
z = \sqrt{- \imath J / \omega_0} e^{-\imath \theta}
\end{equation}
\end{subequations}
and their complex conjugates. Using the same process as for the one-dimensional harmonic oscillator above, we can see that this leads to the correct differential form. By direct substitution into the Hamiltonian, this gives
\begin{equation}
H' = \sqrt{J J^*} \omega_0
\end{equation}
which is a pure function of the actions and is angle-independent. It is interesting to note at this point that the relative phase between the $x$ and $y$ oscillations is actually stored \emph{in the action coordinate} and not in the angle coordinate, and furthermore that there is only one angle variable for a two-dimensional system.

This particular set of coordinates is convenient for harmonic potentials, which can be written as the real or imaginary part of a power series in the complex plane. It greatly simplifies canonical perturbation theory for these potentials, as in the example below. Consider, for example, the Hamiltonian described above with an additional sextupole term:
\begin{equation}
H = \frac{1}{2} \left ( p_x^2 + p_y^2 \right ) + \frac{1}{2} \omega_0^2 \left ( x^2 + y^2 \right ) + \frac{S_3}{3} \left ( x^3 - 3 x y^2 \right )
\end{equation}
In our complex coordinate system, this is simply given as
\begin{equation}
H' = p p^* + \omega_0^2 z z^* + \frac{S'_3}{3} \left [ z^3 + (z^*)^3 \right ]
\end{equation}
where $S'_3$ is rescaled as appropriate. In the action-angle variables, we can write this simply as
\begin{equation}
H' = \sqrt{J J^*} \omega_0 + \frac{S'_3}{3 \omega_0^{3/2}} \left [ J^{3/2} e^{\imath 3 \theta} + (J^*)^{3/2} e^{-\imath 3 \theta} \right ]
\end{equation}
This is algebraically much simpler to work with than the expression in the $x-y$ coordinate system, and a good comparison exercise for the student would be to calculate the first order correction to the angles using perturbation theory in the $x-y$ variables and the $z - z^*$ variables and observe which is more convenient.

\subsection{Harmonically Confined Particle in a Laser Field}

The previous example required the student to work through one canonical change of variables. In a typical application, multiple changes of variable may be needed, and the current method makes these run by much more smoothly. Problems with fast-oscillating and slowly varying coordinates are ubiquitous in plasma and accelerator physics. Examples include various ponderomotive forces from ion traps to laser wakefield acceleration \cite{car_kau_81, taj_daw_79, ros_88}, the free-electron laser instability \cite{ssynima, bpn}, and various forms of electric propulsion \cite{jahn}.

In the previous section I considered a harmonic oscillator. The student can go a step further and look at the problem of a harmonically confined particle in a laser field propagating in the $y$-direction with a vector potential given by
\begin{equation}
\vec{A}_\ell = A_0 \cos \left [ \omega (y/c - t) \right ]\hat{x}
\end{equation}
The Hamiltonian, which is effectively two dimensional, is then given by
\begin{equation}
\begin{split}
H = \frac{1}{2} \left \{ \left ( p_x - \frac{e}{c} A_0 \cos \left [ \omega (y/c - t) \right ] \right )^2 + p_y^2 \right \} + \\ \frac{1}{2} \omega_0^2 \left ( x^2 + y^2 \right )
\end{split}
\end{equation}
This can be solved fairly directly, using multiple canonical transformations. First, expand the Hamiltonian out and write it in terms of the usual $x-y$ action-angle variables derived above to obtain
\begin{equation}
\begin{split}
H' = (J_x + J_y) \omega_0 - \\ \left (\sqrt{J_x \omega_0} \cos \theta_x \right ) \frac{e}{c} A_0 \cos \left [ \omega \left ( \sqrt{J_y/\omega_0} \sin \theta_y - t \right ) \right ]
\end{split}
\end{equation}
where I have assumed that $e A_0/c \ll \sqrt{J_x \omega_0}$ and that higher order term can be ignored. We can rearrange the perturbing term as
\begin{equation}
\begin{split}
\cos \theta_x \cos \left [ \frac{\omega}{c} \left ( \sqrt{J_y/\omega_0} \sin \theta_y - t \right ) \right ] = \\
\frac{1}{2} \biggl \{ \cos \left [ \frac{\omega}{c}  \sqrt{J_y/\omega_0} \sin \theta_y \right ] \left ( \cos (\theta_x - \omega t) + \cos(\theta_x + \omega t) \right ) + \\
\sin \left [ \frac{\omega}{c}  \sqrt{J_y/\omega_0} \sin \theta_y \right ] \left ( \sin (\theta_x + \omega t) - \sin(\theta_x - \omega t) \right ) \biggr \}
\end{split}
\end{equation}
Our first new change of canonical variables is to select the slow oscillations by choosing $\vartheta_x = \theta_x - \omega t$. Then terms that go as $e^{\imath (\vartheta + 2 \omega t)}$ are rapidly oscillating and can be dropped on the average. This is commonly referred to as a transformation to the rotating coordinate system.

To accomplish this canonically, look at the action integral, which in the action-angle variables looks like
\begin{equation}
S = \int J_x d \theta_x + J_y d \theta_y - H' dt
\end{equation}
since we are moving into the slow-varying coordinates, we want
\begin{equation}
J_x d \theta_x + J_y d \theta_y - H' dt = \mathcal{J}_x d \vartheta_x + J_y d \theta_y - \mathcal{H} dt
\end{equation}
and need to solve for $\mathcal{H}$ and $\mathcal{J}_x$. $J_y$ and $\theta_y$ remain unchanged. In the differential form, this criterion reads:
\begin{equation}
J_x d \theta_x  - H' dt = \mathcal{J}_x d \theta_x - \mathcal{H} dt - \mathcal{J}_x \omega dt
\end{equation}
and so $\mathcal{J}_x = J_x$ and $\mathcal{H} = H' - \mathcal{J}_x \omega$. This new, new Hamiltonian is given by
\begin{equation}
\begin{split}
\mathcal{H} = \mathcal{J}_x (\omega_0 - \omega) + J_y \omega_0 + \\ \frac{1}{2} \frac{e}{c} \sqrt{\mathcal{J}_x \omega_0} A_0 \biggl ( \cos \left [ \frac{\omega}{c}  \sqrt{J_y/\omega_0} \sin \theta_y \right ] \cos \vartheta_x - \\ \sin \left [ \frac{\omega}{c}  \sqrt{J_y/\omega_0} \sin \theta_y \right ] \sin \vartheta_x \biggr )
\end{split}
\end{equation}
This is useful because we have used canonical transformations to make the Hamiltonian approximately time-independent in the slow-varying frame. So long as $\omega \gg \omega_0$ this Hamiltonian is conserved, which is a large improvement to the equations of motion. Treating $A_0$ as an expansion parameter, it is then possible to obtain a perturbation expansion in these new canonical variables, which were obtained directly with no intermediate generating function.

\section{Non-Canonical Variables}

Canonical variables are nice because the dynamics are known to be symplectic, which encodes the conservation laws into the dynamics. However, there may come a point where non-canonical variables may be useful. The generating function by design cannot cope with this. However, this method will still yield the correct equations of motion, even if the equations of motion are not symplectic. This method has been used to study magnetohydrodynamics \cite{mor_gre_80, car_lit_82, arn_khe} where the use of Eulerian variables is preferable, even though they are not canonical. Presented here is an example of such variables familiar to most undergraduates.

For the two-dimensional harmonic oscillator, we introduced $z = x + \imath y$ to reduce the complexity of the Hamiltonian. We may do something similar for the one-dimensional case. As an example, consider the one-dimensional harmonic oscillator
\begin{equation}
H = \frac{1}{2} \left ( p^2 + q^2 \right )
\end{equation}
Define the non-canonical variables 
\begin{subequations}
\begin{equation}
a = (p + \imath q)/\sqrt{2}
\end{equation}
\begin{equation}
a^* = (p - \imath q)/\sqrt{2}
\end{equation}
\end{subequations}
This change of variables might be familiar to anyone who took a quantum mechanics course. The Hamiltonian is clearly written as $H = a a^*$. But what are the equations of motion for $a$ and $a^*$? We can turn to the action integral and minimize it using the Euler-Lagrange equations to obtain the non-canonical equations of motion. We know that $p = (a + a^*)/\sqrt{2}$ and $q = - \imath (a - a^*)/\sqrt{2}$ and therefore the action integral is given by
\begin{equation}
\int \left [- \frac{\imath}{2} (a + a^*) \dot{a} + \frac{\imath}{2}(a + a^*) \dot{a}^* - a a^* \right ] ds
\end{equation}
This does not have the same structure of $p dq - H dt$ that the canonical transformations preserve. Specifically, the sign difference between the coefficient of $da$ and $da^*$ prevents the differential above from being written in the canonical form to within a total derivative.

Because the change of variables occurs within the Hamiltonian action integral, we can still obtain equations of motion. Minimizing this action integral using the Euler-Lagrange equation
\[
\delta \int F(x, x') ds = 0 \rightarrow \frac{d}{d s} \frac{\partial F}{\partial x'} - \frac{\partial F}{\partial x} = 0
\]
gives the equations of motion for the non-canonical variables as
\begin{subequations}
\begin{equation}
\dot{a}^* = \imath a
\end{equation}
\begin{equation}
\dot{a} = - \imath a^*
\end{equation}
\end{subequations}

This is identical to the standard equations of motion one might obtain in the Heisenberg picture using these as operators for the quantum mechanical harmonic oscillator. By direct addition of these equations we duplicate the standard equation $\dot{p} = - q$, then by subtraction we obtain that $\dot{q} = p$.

\section{Conclusion}

It is the purpose of this paper to present a short series of examples which exploit the differential form of Hamilton's action integral to find canonical transformations. The generating function is frequently of little direct use aside from finding the new canonical variables, and it is usually the emphasis of conventional classical mechanics textbooks to approach canonical transformations in terms of generating functions. This frequently adds an additional, more complicated step to any calculation that is redundant in almost every case.

By providing a direct derivation of Hamilton's action integral, then working through a series of examples that illustrate the usefulness of this method, it is my hope that this approach to canonical transformations might take root for its practicality and intuitiveness. At the very least, it makes clear where generating functions might come from and how to derive them, although from the point of view described here they are not even strictly necessary.

\section{Acknowledgements}

The author would like to thank Dan T. Abell and David Bruhwiler (Tech-X), Robert Hovden (Cornell U.), Todd Satogata (Jefferson Lab), and Alexander Abanov (Stony Brook University) for helpful discussions.

\bibliography{can_var}

\end{document}